\documentstyle[12pt,psfig]{article}

\newcommand{\rf}[1]{(\ref{#1})}
\newcommand{\bea}{\begin{eqnarray}}
\newcommand{\eea}{\end{eqnarray}}

\newcommand{\e}{\mbox{e}}
\renewcommand{\d}{\mbox{d}}
\newcommand{\g}{\gamma}

\renewcommand{\l}{\lambda}
\renewcommand{\L}{\Lambda}
\renewcommand{\b}{\beta}
\renewcommand{\a}{\alpha}

\newcommand{\ep}{\varepsilon}

\newcommand{\oh}{\frac{1}{2}}

\newcommand{\ra}{\right\rangle}
\newcommand{\la}{\left\langle}
\newcommand{\prt}{\partial}
\newcommand{\mi}{\!-\!}
\newcommand{\equ}{\!=\!}
\newcommand{\pl}{\!+\!}

\newcommand{\tL}{{\tilde{\L}}}
\newcommand{\tX}{{\tilde{X}}}
\newcommand{\tY}{{\tilde{Y}}}

\newcommand{\no}{\nonumber}
\newcommand{\nn}{\no\\}

\newcommand{\ointz}{\oint \frac{dz}{2\pi i \, z}\;}
\newcommand{\ointy}{\oint \frac{dy}{2\pi i \, y}\;}
\newcommand{\ointx}{\oint \frac{dx}{2\pi i \, x}\;}
\newcommand{\SL}{\sqrt{\L}}
\newcommand{\SLT}{\sqrt{\L}T}
\newcommand{\R}{{\rm I\!R}}
\def\void{}
\def\labelmark{}

\newenvironment{formula}[1]{\def\labelname{#1}
\ifx\void\labelname\def\junk{\begin{displaymath}}
\else\def\junk{\begin{equation}\label{\labelname}}\fi\junk}%
{\ifx\void\labelname\def\junk{\end{displaymath}}
\else\def\junk{\end{equation}}\fi\junk\labelmark\def\labelname{}}

{\ifx\void\labelname\def\junk{\end{array}\end{displaymath}}
\else\def\junk{\end{array}\right.\end{equation}}
\fi\junk\labelmark\def\labelname{}\def\junk{}
}

\newcommand{\beq}{\begin{formula}}
\newcommand{\eeq}{\end{formula}}
\newcommand{\beqv}{\begin{formula}{}}

\begin{document}

\hfill NBI-HE-98-11

\hfill AEI-064

\hfill  MPS-RR-98-5

\begin{center}
\vspace{24pt}
{ \large \bf Non-perturbative Lorentzian Quantum Gravity, \\
Causality and Topology Change}

\vspace{24pt}

{\sl J. Ambj\o rn}$\,^{a,}$\footnote{email ambjorn@nbi.dk}
and
{\sl R. Loll}$\,^{b,}$\footnote{email loll@aei-potsdam.mpg.de}

\vspace{24pt}

$^a$~The Niels Bohr Institute, \\
Blegdamsvej 17, DK-2100 Copenhagen \O , Denmark, 

\vspace{24pt}
$^b$~Max-Planck-Institut f\"{u}r Gravitationsphysik,\\
Albert-Einstein-Institut,\\
Schlaatzweg 1, D-14473 Potsdam, Germany

\vspace{36pt}

\end{center}

\vspace{2cm}

\begin{center}
{\bf Abstract}
\end{center}

\vspace{12pt}
\noindent
We formulate a non-perturbative lattice model of two-dimensional 
Lorentzian quantum gravity by performing the path integral 
over geometries with a causal structure. The model can be solved 
exactly at the discretized level. Its continuum limit
coincides with the theory obtained by quantizing 
2d continuum gravity in proper-time gauge, but 
it disagrees with 2d gravity defined via matrix models 
or Liouville theory. By allowing topology change of the compact
spatial slices (i.e.\ baby universe creation), one obtains agreement 
with the matrix models and Liouville theory.


\newpage

\section{Introduction}

A moot point in non-perturbative Euclidean path-integral
approaches to quantum
gravity is the final interpretation of their results in
terms of physical quantities defined in the Lorentzian sector
of the theory. Even classically, it is known that the simple expedient of
applying a Wick rotation $t\rightarrow \tau=it$ fails in all
but a few special cases, for instance, when the space-time is
static and thus admits a {\it global} choice of time $x_0$ such that 
the cross terms $g_{0i}$, $i\geq 1$ of the metric tensor vanish and
its spatial components $g_{ij}$ are time-independent. 
We do not know of a way to set up a 1-to-1 correspondence between
generic solutions of the continuum Einstein equations with different
signatures. If it exists, it is likely to be technically
involved (see, for example, \cite{barbero} for a recent proposal).

From this point of view it is unclear if one 
can expect to obtain the correct Lorentzian theory 
by first performing a path integral over general Euclidean metric 
configurations and then analytically continuing in some 
way. 
Related questions have been discussed in 
\cite{jeff1}, where it was suggested that in the path integral for
gravity with an action in square-root form, and using the rather
unconventional weights $e^{\sqrt{i}S}$, it may be necessary
to sum over both Lorentzian and Euclidean metrics in order to
obtain a unitary evolution. 

To investigate these issues further, we formulate a theory 
of gravity in two space-time dimensions where the summation is restricted 
to metric configurations with a {\em causal} structure, defined on each  
contribution to the sum over states in the 
discrete model under consideration.
In this way one encodes at least part of the Lorentzian 
structure into the Euclidean path integral. In addition to that, we will
consider a suitable analytic continuation of our results.\footnote{
Another way of making 2d gravity more Lorentzian (by allowing for
both time- and spacelike edges in a dynamically triangulated model,
but without causal structures)
was studied in \cite{bj}, but no significantly different behaviour 
from the Euclidean theory was found.} Curiously, our final 
continuum results show some similarity with the approach advocated in 
\cite{jeff1}, although our path-integral construction proceeds along entirely
different lines.

The idea that a notion of causality should be built into each
history that contributes to the path integral amplitude goes
back at least to Teitelboim \cite{teitelboim}, and has more 
recently been advocated in \cite{ms,blms}. However, to our
knowledge a concrete implementation in a well-defined, non-perturbative
model for quantum gravity has so far been missing. An ideal
testing ground for this idea is in two dimensions, where the
Euclidean path-integral construction leads to a non-trivial
gravitational quantum theory. Its properties have been explored
in great detail, both by analytical and numerical methods.
The main reason why this model can be solved analytically, even at the 
discretized level, is that the action in 2d gravity is trivial. 
The Einstein-Hilbert term is a topological invariant and does not 
contribute unless we consider topology changes of space-time.
Moreover, for fixed space-time volume the partition function is 
purely entropic and given by the number of different geo\-metries. 
Exactly this fact is used in the formalism  
of dynamical triangulations (or equivalently, matrix models) to
construct the non-perturbative path integral. One counts the number 
of inequivalent triangulations which can be 
constructed from a given number of triangles of unit volume, 
and with a fixed topology for the resulting simplicial complex.
Inequivalent triangulations (appropriately defined) 
can be related to different geometries \cite{david,adf,adfo,kkm}.
In this way the following quantities have been calculated in pure 
Euclidean 2d gravity: 
\begin{itemize}
\item[(1)]the partition function on the sphere, $Z^{(eu)}(\L)$,
as a function of the renormalized cosmological\cite{david} 
constant $\L$,
\beq{jan1}
Z^{(eu)}(\L) \sim \L^{2-\g} +\ {\rm terms\ less\ singular\ in\ } \L,
~~~~~\g=-\oh,
\eeq
where $\g$ is the so-called string susceptibility exponent;
\item[(2)] the Hartle-Hawking wave functional $W_\L(L)$ 
as a function of the length $L$ of the spatial boundary
(with a marked point) \cite{david1},
\beq{jan2}
W_\L^{(eu)}(L) \propto \frac{1}{L^{5/2}}(1+\SL\, L)\, \e^{-\SL\, L},
\eeq
(more generally, one can calculate ``multi-loop'' correlators 
$W_\L^{(eu)}(L_1,\ldots,L_n)$ \cite{ajm});
\item[(3)]the average Euclidean space-time volume, $B_V(R)$, 
of geodesic balls of radius $R$ in the 
ensemble of universes of Euclidean space-time volume $V$ \cite{aw,ajw,japan},
\beq{jan3}
B_V (R) \propto R^4 F\Bigl( \frac{R^4}{V}\Bigr),
\eeq 
where $F(0)=1$ and $F(x) \propto \e^{-x^{1/3}}$ for $x \to \infty$. Relation 
\rf{jan3} implies that the intrinsic Hausdorff dimension $d_H$ of Euclidean
2d quantum gravity is four (and not two, as one na\"\i vely 
might have expected).\footnote{It can be shown, using only general 
arguments \cite{book1}, 
that Euclidean space-time, with an assumed fractal dimension 
$d_H$, is characterized by a function like \rf{jan3}:
\begin{equation}
B_V(R) \sim R^{d_H}F\Big(R^{d_H}/V\Big), ~~~~F(0)=1,~~~
F(x) = \e^{-x^{1/(d_H-1)}}~~~\mbox{for}~~x\gg 1.
\end{equation}}
\end{itemize}
 
In the following we will show that for a universe with cylinder
topology, restricting the path integral to configurations admitting
a causal structure leads to a theory with $\g= 1/2$ (although the definition
of $\g$ turns out to be ambiguous), a Hartle-Hawking 
wave function with the same exponential decay (in the Euclidean sector)
as \rf{jan2}, but a different functional form, and 
an intrinsic Hausdorff dimension of two, and not four as in \rf{jan3}.
We will further show that once we allow for topology changes of space,
i.e.\ the creation of baby universes, we are led to a theory 
satisfying (1)--(3).

\section{The discrete model}\label{model}

As mentioned in the introduction, the solution of two-dimensional 
quantum gravity
amounts to counting geometries. While this counting problem 
has been solved in Euclidean gravity, it seems non-trivial 
if the space-time has Lorentzian signature. Counting in a field 
theoretical context usually amounts to the introduction of 
a regularization (a discretization) which makes the counting 
procedure well defined. After the counting has been performed 
one may attempt to take the continuum limit of the discretized 
theory. It is unclear which class of geometries to include if 
the signature is not Euclidean. We propose here a model where 
a causal structure is explicitly present in all the geometries 
included in the path integral.

The model is defined as follows. The topology of the underlying manifold
is taken to be $S^1\times [0,1]$, with ``space" represented by the closed 
manifold $S^1$. We consider the evolution of this space in ``time''. 
No topology change of space is allowed at this stage, but we will 
return to this issue in sec.\ \ref{topology}.

\begin{figure}
\centerline{\hbox{\psfig{figure=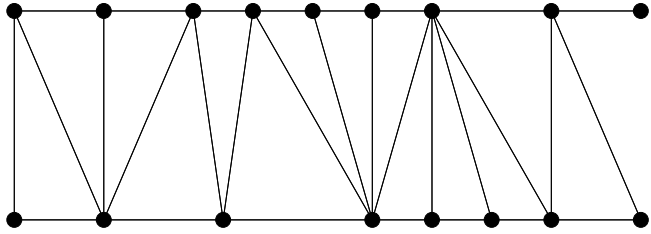,height=3cm,angle=0}}}
\caption[fig1]{The propagation of a spatial slice from step $t$ to 
step $t+1$. The ends of the strip should be joined to form a band
with topology $S^1 \times [0,1]$.}
\label{fig1}
\end{figure}

The geometry of each spatial slice is uniquely characterized by 
the length assigned to it. In the discretized version, the length $L$ 
will be quantized in units of a lattice spacing $a$, i.e.\ 
$L= l\cdot a$ where $l$ is an integer. A slice will thus be 
defined by $l$ vertices and $l$ links connecting them. 
To obtain a 2d geometry, we will evolve this spatial loop in 
discrete steps. This leads to a preferred notion of (discrete) ``time'' 
$t$, where each loop represents a slice of constant $t$.
The propagation from time-slice 
$t$ to time-slice $t+1$ is governed by the following rule: each vertex
$i$ at time $t$ is connected to $k_i$ vertices at time $t+1$, $k_i \geq 1$,
by links which are assigned length $-a$. The $k_i$ vertices, $k_i > 1$, 
at time-slice $t+1$ will be connected by $k_i-1$ consecutive 
space-like links, thus forming $k_i -1$ triangles. 
Finally the right boundary vertex
in the set of $k_i$ vertices will be identified with the left boundary 
vertex of the set of $k_{i+1}$ vertices. In this way we get a total of 
$\sum_{i=1}^l (k_i-1)$ vertices (and also links) at time-slice $t+1$ and 
the two spatial slices are connected by $\sum_{i=1}^l k_i
\equiv l_{t}+l_{t+1}$ triangles.
See fig.\ \ref{fig1}. 

The elementary building blocks of a geometry are therefore triangles
with one space- and two time-like edges. We define them to be flat
in the interior. A consistent way of assigning interior angles to
such Minkowskian triangles is described in \cite{sorkin}. The
angle between two time-like edges is $\gamma_{tt}=-\arccos \frac{3}{2}$,
and between a space- and a time-like edge $\gamma_{st}=
\frac{\pi}{2}+\frac{1}{2} \arccos \frac{3}{2}$, summing up to
$\gamma_{tt}+2\gamma_{st}=\pi$. The sum over all angles around a
vertex with $j$ incoming and $k$ outgoing time-like edges (by
definition $j,k\geq 1$) is given by $2\pi+(4-j-k)\arccos\frac{3}{2}$.
The regular triangulation of flat Minkowski space
corresponds to $j=k=2$ at all vertices. The volume of a single
triangle is given by $\frac{\sqrt{5}}{4}a^2$.

One may view these geometries as a subclass of all possible
triangulations that allow for the introduction of a causal
structure. Namely, if we think of all time-like links as being
future-directed, a vertex $v'$ lies in the future of a
vertex $v$ iff there is an oriented sequence of time-like
links leading from $v$ to $v'$. Two arbitrary vertices may or may not
be causally related in this way. 

In quantum gravity we are instructed to sum over all geometries connecting, 
say, two spatial boundaries of length $L_1$ and $L_2$, with the weight 
of each geometry $g$ given by 
\beq{1}
\e^{i S[g]}, ~~~~~S[g] = \L \int \!\!\sqrt{-g}~~~(\mbox{in 2d}),
\eeq
where $\L$ is the cosmological constant.
In our discretized model the boundaries will be characterized by 
integers $l_1$ and $l_2$, the number of vertices or links at the two
boundaries. The path integral amplitude for the propagation from 
geometry $l_1$ to $l_2$ will be the sum over all interpolating 
surfaces of the 
kind described above, with a weight given by the discretized version of 
\rf{1}. Let us call the corresponding amplitude $G^{(1)}_\l(l_1,l_2)$.
Thus we have
\bea
G_\l^{(1)}(l_1,l_2) &=& \sum_{t=1}^{\infty} G_\l^{(1)}(l_1,l_2;t),\label{3}\\
G_\l^{(1)}(l_1,l_2;t) &=& 
\sum_{l=1}^\infty G_\l^{(1)}(l_1,l;1)\;l\;G_\l^{(1)}(l,l_2,t-1),\label{4}\\
G_\l^{(1)}(l_1,l_2;1) &=& \frac{1}{l_1}\sum_{\{k_1,\dots,k_{l_1}\}} 
\e^{i \l a^2 \sum_{i=1}^{l_1} k_i}, \label{5}
\eea
where $\l$ denotes the {\em bare}  
cosmological constant\footnote{One obtains 
the renormalized (continuum) cosmological constant $\L$ in \rf{1} by 
an additive renormalization, see below.} (we have absorbed the finite
triangle volume factor), and where $t$ denotes the 
total number of time-slices connecting $l_1$ and $l_2$. 

From a combinatorial point of view it is convenient to mark a 
vertex on the entrance loop in order to get rid of the factors
$l$ and $1/l$ in \rf{4} and \rf{5}, that is,
\beq{6}
G_\l (l_1,l_2;t) \equiv l_1 G_\l^{(1)}(l_1,l_2;t)
\eeq
(the unmarking of a point may be thought of as 
the factoring out by (discrete) spatial diffeomorphisms).
Note that $G_\l(l_1,l_2;1)$ plays the role of a 
transfer matrix, satisfying
\bea
G_\l(l_1,l_2,t_1+t_2) &=& \sum_{l} G_\l(l_1,l;t_1)\; G_\l(l,l_2;t_2)\label{7}\\
G_\l(l_1,l_2;t+1) &=& \sum_{l} G_{\l}(l_1,l;1)\;G_\l(l,l_2;t).\label{8}
\eea
Knowing $G_\l(l_1,l_2;1)$ allows us to find $G_\l(l_1,l_2;t)$
by iterating \rf{8} $t$ times. This program is conveniently 
carried out by introducing the generating function for the numbers
$G_\l(l_1,l_2;t)$,
\beq{9}
G_\l(x,y;t)\equiv \sum_{k,l} x^k\,y^l \;G_\l(k,l;t),
\eeq
which we can use to rewrite \rf{7} as 
\beq{10}
G_\l(x,y;t_1+t_2) = \ointz G_\l(x,z^{-1};t_1) G_\l(z,y;t_2),
\eeq
where the contour should be chosen to include the singularities 
in the complex $z$--plane of $G_\l(x,z^{-1};t_1)$ but not those
of $G_\l(z,y;t_2)$. 

One can either view the introduction of $G_\l(x,y;t)$ as a purely
technical device or take $x$ and $y$ as boundary cosmological 
constants,
\beq{10a}
x=\e^{i\l_ia},~~~~y=\e^{i\l_oa},
\eeq
such that $x^k= \e^{i\l_i a\,k}$ becomes a boundary cosmological term,
and similarly for $y^l= \e^{i\l_o a\, l}$. 
Let us for notational convenience define 
\beq{11}
g=\e^{i\l a^2}.
\eeq
For the technical purpose of counting we view $x,y$ and $g$ as 
variables in the complex plane. In general the function 
\beq{11a}
G(x,y;g;t)\equiv G_\l(x,y;t)
\eeq
will be analytic in a neighbourhood of $(x,y,g)=(0,0,0)$.  

From the definitions \rf{5} and \rf{6} it follows by standard techniques 
of generating functions that we may associate a factor $g$ with each 
triangle, a factor $x$ with each vertex on the entrance loop and 
a factor $y$ with each vertex on the exit loop, leading to
\beq{12}
G(x,y;g;1) =\sum_{k=0}^\infty \left( gx \sum_{l=0}^\infty
 (gy)^l \right)^k -
\sum_{k=0}^\infty (gx)^k = \frac{g^2 xy}{(1-gx)(1-gx-gy)}.
\eeq
Formula \rf{12} is simply a book-keeping device for all possible
ways of evolving from an entrance loop of any length in one step to
an exit loop of any length. The subtraction of the term $1/(1-gx)$ 
has been performed to 
exclude the degenerate cases where either the entrance or the exit 
loop is of length zero. 
  
From \rf{12} and eq.\ \rf{10}, with $t_1=1$, we obtain
\beq{13}
G(x,y;g;t) = \frac{gx}{1-gx}\; G(\frac{g}{1-gx},y;g;t-1).
\eeq
This equation can be iterated and the solution written as 
\beq{14}
G(x,y;g;t) = F_1^2(x)F_2^2(x) \cdots F_{t-1}^2(x) 
\frac{g^2 xy}{[1-gF_{t-1}(x)][1-gF_{t-1}(x)-gy]},
\eeq
where $F_t(x)$ is defined iteratively by
\beq{15}
F_t(x) = \frac{g}{1-gF_{t-1}(x)},~~~F_0(x)=x.
\eeq
Let $F$ denote the fixed point of this iterative equation. By standard
techniques one readily obtains
\beq{16}
F_t(x)= F\ \frac{1-xF +F^{2t-1}(x-F)}{1-xF +F^{2t+1}(x-F)},~~~~
F=\frac{1-\sqrt{1-4g^2}}{2g}.
\eeq
Inserting \rf{16} in eq.\ \rf{14}, we can write
\bea
G(x,y;g,t)\!\!\! &=&\!\!\!\!  \frac{ F^{2t}(1-F^2)^2\; xy}
{(A_t-B_tx)(A_t-B_t(x+y)+C_txy)}
\label{17}\\
~\!\!\!& =&\!\!\!\!
 \frac{F^{2t}(1-F^2)^2\;xy}{\Big[(1\!\!-\!xF)\!-\!F^{2t+1}(F\!\!-\!x)\Big]
\Big[(1\!\!-\!xF)(1\!\!-\!yF)\!-\!F^{2t} (F\!\!-\!x)(F\!\!-\!y)\Big]}~,
~~~\label{17a}
\eea
where the time-dependent coefficients are given by 
\beq{18}
A_t =1-F^{2t+2},~~~B_t=F(1-F^{2t}),~~~C_t=F^2(1-F^{2t-2}).
\eeq
The combined region of convergence to the 
expansion in powers $g^kx^ly^m$, valid for all $t$ is 
\beq{18a}
|g| < \oh,~~~~ |x|< 1,~~~~|y|<1.
\eeq
The asymmetry between $x$ and $y$ in the expressions \rf{17} and \rf{17a}
is due to the marking of the entrance loop. If we also mark the exit loop
we have to multiply $G_\l(l_1,l_2;t)$ by $l_2$. We define
\beq{18b}
G_\l^{(2)}(l_1,l_2;t) \equiv l_2\, G_\l(l_1,l_2;t)= l_1l_2 
G^{(1)}_\l (l_1,l_2;t).
\eeq
The corresponding generating function $G^{(2)}(x,y;g;t)$ is obtained from
$G(x,y;g;t)$ by acting with $y \frac{d}{dy}$,
\beq{18c}
G^{(2)}(x,y;g;t) = 
\frac{ F^{2t}(1-F^2)^2 \, xy}{(A_t -B_t(x+y)+C_t xy)^2}.
\eeq

We can compute $G_\l(l_1,l_2;t)$ from $G(x,y;g;t)$ by
a (discrete) inverse Laplace transformation
\beq{19}
G_\l(l_1,l_2;t) =\ointx \ointy \frac{1}{x^{l_1}}\, 
\frac{1}{y^{l_2}}\; G(x,y;g;t),
\eeq
where the contours should be chosen in the region where $G(x,y;g;t)$ is 
analytic. A more straightforward method is to rewrite the
right-hand side of \rf{17} as a power series in $x$ and $y$,
yielding
\beq{19aNEW!}
G_\l (l_1,l_2;t) = 
\frac{F^{2t}(1-F^2)^2 B^{l_1\pl l_2}}{l_2\;\; A^{l_1\pl l_2\pl 2}} \;
\;\sum_{k=0}^{\min (l_1,l_2)\mi 1} 
\frac{l_1\pl l_2\mi k\mi 1}{k!(l_1\mi k\mi 1)!(l_2\mi k\mi 1)!}
\left( \mi \frac{A_tC_t}{B_t^2}\right)^k.
\eeq
which, as expected, is symmetric with respect to $l_{1}$ and $l_{2}$
after division by $l_{1}$. 

In the next section we will give explicit expressions  
for $G_\l(l_1,l_2;t)$, $G_\l(l_1,l_2)$ and $G_\l(x,y)$ 
(the integral of $G_\l(x,y;t)$ over $t$) in a certain continuum limit.

\section{The continuum limit}

The path integral formalism we are using here
is very similar to the one used to re\-pre\-sent the free particle as 
a sum over paths. Also there one performs a
summation over geometric objects (the paths), and the path integral itself
serves as the propagator. From the particle case it is known that the bare mass
undergoes an additive renormalization (even for the free particle), 
and that the bare propagator is subject to a wave-function renormalization
(see \cite{book1} for a review). The same is true
in two-dimensional gravity, treated in the formalism of 
dynamical triangulations \cite{book1}. The coupling constants
with positive mass dimension, i.e.\ the cosmological constant and the 
boundary cosmological constants, undergo an 
additive renormalization, while the partition function itself (i.e.\ the 
Hartle-Hawking-like wave functions) 
undergoes a multiplicative wave-function renormalization.
We therefore expect the bare coupling constants $\l,\l_i$ and 
$\l_0$ to behave as 
\beq{20a}
\l = \frac{C_{\l}}{a^2} + \tilde{\L},~~~~
\l_i= \frac{C_{\l_{i}}}{a}+\tilde{X},~~~
\l_o =\frac{C_{\l_{o}}}{a}+\tilde{Y},
\eeq
where $\tilde{\L},\tilde{X},\tilde{Y}$ denote the renormalized 
cosmological and boundary cosmological constants. If we introduce
the notation 
\beq{20c}
g_c = \e^{i C_{\l}},~~~~x_c= \e^{i C_{\l_{i}}},~~~~y_c=\e^{iC_{\l_{o}}},
\eeq
for critical values of the coupling constants, 
it follows from \rf{10a} and \rf{11} that 
\beq{20b}
g=g_c\,\e^{ia^2\tL},~~~~x=x_c\,\e^{ia\tX},~~~~y=y_c\,\e^{ia\tY}.
\eeq 
The wave-function renormalization will appear as a multiplicative  
cut-off dependent factor in front of the bare 
``Green's function'' $G(x,y;g;t)$, 
\beq{20}
G_\tL (\tX,\tY;T) = \lim_{a \to 0} a^{\eta} G(x,y;g;t),
\eeq
where $T=a\, t$, and where the critical exponent $\eta$ 
should be chosen so that 
the right-hand side of eq.\ \rf{20} exists. In general this will only be 
possible for particular
choices of $g_c,x_c$ and $y_c$ in \rf{20}. 

The basic relation \rf{7} can survive the limit \rf{20} only 
if $\eta=1$, since we have assumed that the boundary lengths 
$L_1$ and $L_2$ have canonical dimensions and satisfy $L_i = a\, l_i$. 

From eqs.\ \rf{17} and \rf{18} it is clear that we can only 
obtain a non-trivial continuum limit if $|F| \to 1$.
This leads to a one-parameter family of possible choices
\beq{23}
g_c= \frac{1}{2\cos \a}~~~~\mbox{for}~~~F=\e^{i\a},~~~ \a\in \R,
\eeq
for critical values of $g$. 
It follows from \rf{11} that most values of $g_c$ correspond 
to a complex {\em bare} cosmological constant $\l$. However, the 
renormalized cosmological constant $\tilde{\L}$ in 
\rf{20a} (depending on how we approach $g_c$ in the complex plane)
could in principle still be real.  

\begin{figure}
\centerline{\hbox{\psfig{figure=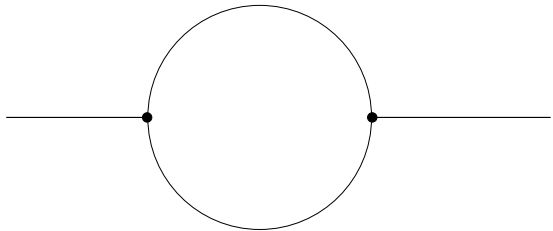,height=3cm,angle=0}}}
\caption[fig2]{The circle of convergence in the complex $g$ plane (radius 1/2),
and the critical lines, ending in $g=\pm 1/2$.}
\label{fig2}
\end{figure}

A closer analysis reveals that only at $g_c=\pm 1/2$, 
corresponding to $\a=0,\pi $, is there 
any possibility of obtaining an interesting continuum limit.
Note that these two values are the only ones  
which can be reached from a region of 
convergence of $G(x,y;g;t)$ (see fig.\ \ref{fig2}).
Note also that requiring the bare $\lambda$ to lie inside 
the region of convergence when $g \to g_c$ leads to a restriction 
Im$\, \tL > 0$ on the {\em renormalized} cosmological constant $\tL$,
since $|g|<\frac{1}{2}\Rightarrow \mbox{Im}\ \lambda >\ln 2$.

Without loss of generality, we will consider the critical value
$g_c=1/2$. It corresponds to 
a purely imaginary bare cosmological constant
$\l_{c}:=C_{\l}/a^{2} = -i \ln 2/a^2$.
If we want to approach this point from the region in the 
complex $g$-plane where $G(x,y;g;t)$  
converges it is natural to choose the renormalized coupling $\tL$
imaginary as well, $\tL = i\L$, i.e.
\beq{25a}
\l = i\ \frac{\ln \oh}{a^2} +i \L.
\eeq
One obtains a well-defined scaling limit (corresponding to
$\L\in \R$) by letting $\l\to \l_{c}$ along the imaginary axis.
The Lorentzian form for the continuum 
propagator is obtained by an analytic continuation $\L\to -i\L$
in the {\it renormalized} coupling of the resulting Euclidean 
expressions. 

At this stage it may seem that we are surreptitiously reverting
to a fully Euclidean model. We could of course equivalently 
have conducted
the entire discussion up to this point in the ``Euclidean sector'',
by omitting the factor of $-i$ in the exponential \rf{1} of the
action, choosing $\l$ positive real and taking all edge lengths 
equal to 1. However, from a purely Euclidean point of view there
would not have been any reason for restricting the state sum to a subclass
of geometries admitting a causal structure. The associated 
preferred notion of a discrete ``time'' allows us to define
an ``analytic continuation in time''. Because of the simple
form of the action in two dimensions, the rotation 
\beq{25b}
\int  dx\ dt \sqrt{-g_{lor}} \to i\int  dx \ dt_{eu} \sqrt{g_{eu}}
\eeq
to Euclidean metrics in our model is equivalent to the analytic continuation
of the cosmological constant $\L$.

From \rf{17} or \rf{17a} it 
follows that we can only get macroscopic loops in the limit 
$a \to 0$ if we simultaneously take $x,y \to 1$. (For $g_c=-1/2$, one
needs to take $x,y \to -1$. The continuum expressions one obtains
are identical to those for $g_c=1/2$.) Again 
the critical points correspond to purely imaginary 
bare boundary cosmological coupling constants. We will 
allow for such imaginary couplings and thus approach the 
critical point $\l_i= \l_o=0$ from the region of convergence of 
$G(x,y;g;t)$, i.e.\ via real, positive $X,Y$ where  
\beq{25c}
\l_i = i X a,~~~~\l_o=i Y a.
\eeq
Again $X$ and $Y$ have an obvious interpretation as positive boundary
cosmological constants in a Euclidean theory, which may be
analytically continued to imaginary values to reach the Lorentzian
sector.

Summarizing, we have 
\beq{25}
g=\oh \e^{-\L a^2} \to \oh (1-\oh \L a^2),~~~(\mbox{i.e.}~~F=1-a\sqrt{\L})
\eeq
as well as 
\beq{25d}
x=\e^{-Xa} \to 1-aX,~~~~~~y=\e^{-aY} \to 1-aY,
\eeq
where the arrows $\to$ in \rf{25} and \rf{25d} should be viewed 
as analytic coupling constant redefinitions of $\L,X$ and $Y$,
which we have performed to get rid of factors of 1/2 etc. in the formulas
below.
With the definitions \rf{25} and \rf{25d} it is straightforward 
to perform the continuum limit of $G(x,y;g,t)$ as $(x,y,g) \to 
(x_c,y_c,g_c)=(1,1,1/2)$, yielding
\bea
G_\L(X,Y;T) &=& \frac{4\L\ \e^{-2\SLT}}{(\SL+X)+\e^{-2\SLT}(\SL-X)}\nonumber\\
&&\times \, \frac{1}{(\SL+X)(\SL+Y)-\e^{-2\SLT}(\SL-X)(\SL-Y)}.
\label{26}
\eea
For $T \to \infty$ one finds 
\beq{27}
G_\L(X,Y;T) \buildrel{T\rightarrow
\infty}\over\longrightarrow \frac{4\L \;\e^{-2\SLT}}{(X+\SL)^2(Y+\SL)}.
\eeq

From $G_\L(X,Y;T)$
we can finally calculate $G_\L(L_1,L_2;T)$, the continuum 
amplitude for propagation from a loop of length $L_1$, 
with one marked point, at time-slice $T=0$ to a loop of length $L_2$ 
at time-slice $T$, by an inverse Laplace transformation,
\beq{22}
G_\L(L_1,L_2;T) = \int_{-i\infty}^{i\infty} d X \int_{-i\infty}^{i\infty} d Y
\; \e^{X L_1}\;\e^{Y L_2}\; G_\L(X,Y;T).
\eeq
This transformation can be viewed as the limit of \rf{19} for 
$a \to 0$. The continuum version of \rf{10} thus reads
\beq{22a}
G_\L(X,Y;T_1+T_2) = \int_{-i\infty}^{i\infty}d Z \; 
G_\L(X,-Z;T_1) \, \,G_\L(Z,Y;T_2),
\eeq
where it is understood that the complex contour of integration 
should be chosen to the left of 
singularities of $G_\L(X,-Z;T_{1})$, but to the right of those of 
$G_\L(Z,Y,T_{2})$. 
By an inverse Laplace transformation we get in the limit
$T\rightarrow\infty$
\beq{27a}
G_\L(L_1,L_2;T) \buildrel{T\rightarrow
\infty}\over\longrightarrow  4 L_1 \e^{-\SL (L_1+L_2)} \;\e^{-2\SLT},
\eeq
where the origin of the factor $L_1$ is the marking of a 
point in the entrance loop. For $T\to 0$ we obtain
\beq{28}
G_\L(X,Y;T) \buildrel{T\rightarrow
0}\over\longrightarrow \frac{1}{X+Y},
\eeq
in agreement with the expectation that the inverse Laplace transform
should behave like
\beq{29}
G_\L(L_1,L_2;T) \buildrel{T\rightarrow
0}\over\longrightarrow \delta(L_1-L_2).
\eeq
The general expression for $G_\L(L_1,L_2;T)$ can be computed
as the inverse Laplace transform 
of formula \rf{26}, yielding
\beq{30}
G_\L(L_1,L_2;T) = \frac{\e^{-[\coth \SLT] \SL(L_1+L_2)}}{\sinh \SLT}
\; \frac{\sqrt{\L L_1 L_2}}{L_2}\; \; 
I_1\left(\frac{2\sqrt{\L L_1 L_2}}{\sinh \SLT}\right), 
\eeq
where $I_1(x)$ is a modified Bessel function of the first kind.
The asymmetry between $L_1$ and $L_2$ arises because the entrance loop 
has a marked point, whereas the exit loop has not. The amplitude with 
both loops marked is obtained by multiplying with $L_2$, while the 
amplitude with no marked loops is obtained after dividing 
\rf{30} by $L_1$. Quite remarkably, our highly non-trivial 
expression \rf{30} agrees
with the loop propagator obtained from a bona-fide continuum calculation
in proper-time gauge of pure 2d gravity by Nakayama \cite{nakayama}.
More precisely, his propagator $A_m$ for $m=0$ ($m$ is a winding number
introduced in a somewhat ad-hoc manner in \cite{nakayama}) is related
to ours by $G_\L(L_1,L_2;T)=\frac{L_1}{L_2} A_0(L_1,L_2;T)$, which just 
reflects the fact that the exit instead of the entrance loop has been marked.
Also additional ambiguities in the continuum formulation, 
involving shifts $m \to m+1/2$ due to renormalization, are 
fixed in our approach. 

To obtain the propagator of the Lorentzian theory, we substitute
$\L\to -i\L$ in \rf{30}. As a consequence, the amplitude becomes
complex and the hyperbolic functions pick up oscillatory contributions. 
Both the real and the imaginary parts continue to be exponentially
damped for large $T$. 
What is at first puzzling about the functional form of \rf{30} is
that the na\"\i ve analytical continuation in ``time'', $T\to -i T$,
leads to a drastically different (and highly singular) result. 
However, this is an incorrect choice, which can be understood as 
follows. The combination $\SL T$ appearing as arguments in \rf{30}
arises in taking the continuum limit of powers of the form $F^{t}$
in expressions like \rf{17}, \rf{17a}, where $F$ is defined in \rf{16}.

There are two aspects to a possible analytic continuation of 
$F^t$. The power $t$ in $F^t$ should clearly not be continued,
since it is simply an integer counting
the number of iterations of the transfer matrix. 
However, the function $F$ itself does refer
to the action, because the dimensionless coupling constant
$g = \e^{i\l a_t a_l}$ is the action for a single Lorentzian triangle. 
(For added clarity we have distinguished between the lattice spacings
in time- and space-directions, and called them $a_{t}$ and $a_{l}$.)
From the expression for $F$ in terms of $g$ in \rf{16}, we have
$F=1-\sqrt{a_{t}a_{l}\L}$.
The analytic continuation of $F$ in time, from Euclidean to Lorentzian 
time, corresponds to the substitution $a_{t}\to -i\ a_{t}$ 
under the square-root sign, and thus becomes equivalent 
to the continuation $\L \to -i\L$ in the cosmological constant, 
as already remarked below eq.\ \rf{25b}. 
The subtleties associated with the 
analytical continuation in the ``time''-parameter $T$ appearing in a 
transfer-matrix formulation of quantum gravity 
were first discussed in \cite{jeff1,jeff2} in the context of a
square-root action formulation.
They will be present also in more complicated theories,
where the analytic continuation from Euclidean metrics to 
Lorentzian metrics cannot be absorbed by
a similar continuation in $\L$ \cite{jeff1}.

Finally, we compute the amplitude describing the transition from 
$L_1$ to $L_2$ for an arbitrary ``time''-separation of the slices
by integrating over $T$. From \rf{30}, multiplied by $L_2$
in order to arrive at the symmetric propagator where both loops
are marked, we get
\beq{31}
G^{(2)}_\L(L_1,L_2) = 
\int_0^\infty dT\ G^{(2)}_\L(L_1,L_2;T)=
\frac{\e^{-\SL | L_1-L_2|}- \e^{-\SL(L_1+L_2)}}{2\SL }. 
\eeq
One of course obtains the same result by first integrating
$G_\L(X,Y;T)$ with respect to $T$, and then doing the inverse
Laplace transform. Again the analytic continuation $\L \to -i\L$
leads to a complex amplitude.

\section{The differential equation}

The basic result \rf{26} for $G_\L(X,Y;T)$
can be derived by taking the continuum limit of 
the recursion relation  \rf{13}. By inserting \rf{25} and \rf{25d} 
in eq.\ \rf{13}
and expanding to first order in the lattice spacing $a$ we obtain
\beq{32} 
\frac{\prt}{\prt T} G_\L(X,Y;T) + \frac{\prt}{\prt X}
\Bigl[ (X^2-\L) G_\L(X,Y;T) \Bigr]=0.
\eeq
This is a standard first order partial differential equation which 
should be solved with the boundary condition \rf{28} at $T=0$, since this
expresses the natural condition \rf{29} on $G_\L(L_1,L_2)$.
The solution is thus 
\beq{33}
G_\L(X,Y;T) = \frac{\bar{X}^2(T;X)-\L}{X^2-\L}\; \frac{1}{\bar{X}(T;X)+Y}, 
\eeq
where $\bar{X}(T;X)$ is the solution to the characteristic equation
\beq{34}
\frac{d \bar{X}}{dT} = -(\bar{X}^2-\L),~~~~\bar{X}(T=0)=X.
\eeq
It is readily seen that the solution is indeed given by \rf{26}
since we obtain
\beq{35}
\bar{X}(T) = \SL \; 
\frac{(\SL+X)-\e^{-2\SLT}(\SL-X)}{(\SL+X)+\e^{-2\SLT}(\SL-X)}.
\eeq

If we interpret the propagator $G_\L(L_1,L_2;T)$ as the matrix element
between two boundary states of a Hamiltonian evolution in 
``time'' $T$,
\beq{ham}
G_\L(L_1,L_2;T)=<L_1|\e^{-\hat H T}|L_2>
\eeq 
we can, after an inverse Laplace transformation, read off the functional form
of the Hamiltonian operator $\hat H$ from \rf{32},
\beq{35b}
\hat H(L,\frac{\partial}{\partial L})=
 -\frac{\partial^2}{\partial L^2}+\L .
\eeq
Using \rf{31}, it is now straightforward to check that 
\beq{35c}
\hat H(L_1,\frac{\partial}{\partial L_1})\ G^{(2)}_\L(L_1,L_2)=
 \delta (L_1-L_2),
\eeq
as is expected for the propagator. 
The corresponding Hamiltonian for the propagator of unmarked loops
is given by
\beq{35e}
\hat H_u(L,\frac{\partial}{\partial L})=
-L \frac{\partial^2}{\partial L^2}-2 \frac{\partial}{\partial L}
+\L L.
\eeq

A solution to the ``Wheeler-DeWitt equation" is only obtained if we 
integrate the expression \rf{30} over the entire $T$-axis (as observed
a long time ago in \cite{teitelboim}):
\beq{35d}
\hat H(L_1,\frac{\partial}{\partial L_1})
\int_{-\infty}^\infty dT\ G^{(2)}_\L(L_1,L_2;T)=
\hat H(L_1,\frac{\partial}{\partial L_1})
\ \frac{\sinh \sqrt{\L} (L_1+L_2)}{\sqrt{\L}}=0.
\eeq

The above construction refers to the evolution of the system
with respect to the ``time''-parameter $T$ appearing in the
transfer-matrix approach. However, we have argued earlier
that one should {\it not} simply analytically continue $T \to -iT$ to relate
the Euclidean and Lorentzian sectors of the theory. 
Not taking $T$ seriously as a time-parameter 
presumably implies that also the operator $\hat H$ appearing in \rf{ham}
is not the physically relevant Hamiltonian. 

At any rate, our choice of analytic continuation does not seem to lead to
a self-adjoint Hamiltonian if one uses the prescription \rf{ham} 
for the Lorentzian case. A possible way out 
may be to use weights of the form $\e^{\sqrt{i}S}$ {\em and} to sum 
over a class of both Lorentzian and Euclidean 
geometries, as advocated in \cite{jeff1}. Like in this approach, we are also 
encountering the factor $\sqrt{i}$, 
but it is presently unclear to us whether the two can be related.  
 

\section{Observables}

We will now compare the predictions coming from the Euclidean sector
of our model with those from 2d quantum gravity as defined
by matrix models (or Liouville theory). As mentioned in the introduction,  
this amounts to calculating the string susceptibility exponent $\g$,
the Hartle-Hawking wave function $W_\L(L)$ and the intrinsic Hausdorff
dimension $d_H$. 
We must first {\em define} what we mean by the disc amplitude 
$W_\L(L)$ in our model. A natural definition is given by
\beq{extra1}
W_\L (L) :=  G_\L(L,L_2\equ 0) = \e^{-\SL L},
\eeq
where the last equality follows from \rf{31}. We have contracted
the exit loop to length zero in order to produce a disc.
This procedure leaves a ``mark'' at the very endpoint of the universe,
contrary to the usual definition of $W_\L(L)$ in Euclidean 2d quantum 
gravity. Had we applied a definition analogous to \rf{extra1}
in Euclidean gravity, setting $L_{2}=0$ in the propagator would
also have led to a marking inside the disc. To relate it to the
usual disc amplitude $W_\L^{(eu)}(L)$, formula \rf{extra1} would 
have had to be replaced by
\beq{extra2}
-\frac{\partial}{\partial \L} W_\L^{(eu)}(L) 
= G^{(eu)}_{\L}(L,L_2\equ 0) ~~(\propto \frac{1}{\sqrt{L}}\, \e^{-\SL L}).
\eeq
The derivative appears because marking a point in Euclidean gravity 
is equivalent to differentiating with respect to the cosmological constant. 
However, in our model the mark in \rf{extra1} does not correspond to 
a differentiation $\partial /\partial\L$, since unlike for Euclidean
gravity the mark cannot be located anywhere in the bulk. 
In fact, from the definition of 
$G_\L(L_1,L_2;T)$ it follows that 
\beq{extra3}
W_\L(L) = \int_0^\infty \d T \; G_\L(L, L_2 \equ 0;T),
\eeq 
which implies that the mark is always at the latest proper time $T$.
Although quite similar, \rf{extra1} differs from \rf{extra2}, a fact which 
becomes more obvious when we consider their respective Laplace transforms,
whose regular parts are given by 
\beq{ob3}
W_\L(X) = \frac{1}{X+\SL},~~~~~
W_\L^{(eu)}(X) = (X-\oh \SL) \sqrt{X+\SL}.
\eeq 

From $W_\L^{(eu)}(L)$ we can extract $\g$ as follows: 
by its very definition, $\g$ is the critical 
exponent controlling the leading {\em non-analytical} behaviour
of the partition function $Z^{(eu)}(\L)$ as a function of the 
cosmological constant $\L$.
From definition \rf{jan1} and the arguments given above, differentiating
$Z^{(eu)}(\L)$ twice with respect to $\L$ leads to the partition 
function where we sum over closed 
surfaces of spherical topology with two marked points.
In \rf{extra2} we have already marked 
one point on the disc by differentiating with respect to $\L$. 
All that remains to be done 
in order to create closed surfaces with two marked points
is to divide $W^{(eu)}_\L (L)$ by $L$ to remove a factor
proportional to the boundary length $L$ (which was originally introduced 
for combinatorial convenience), and to contract the boundary loop $L$ to zero.
This gives 
\beq{ob4}
-\frac{\partial}{\partial \L}\; \frac{W_\L^{(eu)}(L)}{L} = \frac{1}{L^{3/2}}-
\frac{\SL}{L^{1/2}} + \cdots.   
\eeq
Expanding this expression in $L$ in the limit
$L\to 0$, and extracting the first non-analytic power of $\L$ leads
to the critical exponent  
\beq{ob5}
\g= -\oh.
\eeq
Following the analogous procedure for the disc amplitude \rf{extra1} leads 
to $\g=1/2$.

However, the physical interpretation of $\g$ in our model cannot be
the same as in Liouville gravity. Firstly, the standard interpretation
of $\g=1/2$ would be that we are dealing with objects 
with the fractal structure of so-called branched polymers \cite{adj},
which is obviously not the case. 
We also remind the reader that apart from characterizing 
the leading singularity in the partition function, in Euclidean 
quantum gravity the exponent $\g$ 
governs the rate of baby universe creation \cite{jain,jain1,jain2}.
This is implicit in the expansion \rf{ob4} in that $\SL$ is multiplied 
by a divergent power of $L$. This singularity reflects the 
proliferation of baby universes at the cut-off scale. By contrast,
the analogous term in the expansion 
\beq{ob6}
\frac{W_\L(L)}{L} = \frac{1}{L}-\SL+ \cdots   
\eeq
for \rf{extra1} is non-singular, in correspondence with the fact that 
by construction our model contains no baby universes.

The proliferation of baby universes is closely related to 
the intrinsic Hausdorff dimension $d_H=4$ in Euclidean 
2d quantum gravity, as decribed by the relation \rf{jan3}. 
As a consequence of 
this fractal dimensionality, the geodesic distance $R$, or equivalently
the ``time $T$'' in the amplitude $G^{(eu)}_\L(L_1,L_2;T)$ 
has anomalous length dimension 
$[{\rm L}]^{1/2}$. This also implies that for large $T$ the average 
``spatial'' volume of a slice at some intermediate time will
have an anomalous dimension. In fact, in Euclidean 2d quantum gravity  
we have 
\beq{ob7}
G^{(eu)}_\L(L_1,L_2;T) \propto \e^{-\sqrt[4]{\L} T}~~~~\mbox{for}~~~
T \to \infty,
\eeq
from which one can calculate the average {\em two-dimensional} 
volume $V(T)$ in 
the ensemble of universes with two boundaries separated by a
geodesic distance $T$,
\beq{ob8}
\la V(T) \ra = - \frac{1}{G^{(eu)}_\L(L_1,L_2;T)} 
\frac{\partial}{\partial \L}\;
G^{(eu)}_\L(L_1,L_2;T) \propto \frac{T}{\L^{3/4}}.
\eeq
For large $T$ we therefore expect the average spatial volume $L_{space}$ 
at intermediate $T$'s to behave like
\beq{ob9}
\la L_{space} \ra = \frac{\la V(T) \ra}{T} \propto \frac{1}{\L^{3/4}}.
\eeq
In the present model, according to \rf{27a}, the amplitude
behaves for large $T$ like 
\beq{ob10}
G_\L(L_1,L_2;T) \propto \e^{-\SL T},
\eeq
which simply means that the dimension of $T$ in this case is 
$[{\rm L}]$. We therefore obtain instead of \rf{ob9} the dependence
\beq{ob11}
\la L_{space} \ra \propto \frac{1}{\SL}.
\eeq
This reflects the fact that the quantum space-time of our model does 
not have an anomalous fractal dimension, and thus differs drastically 
from the average space-time in the usual two-dimensional Euclidean quantum 
gravity.

It is possible to calculate explicitly $\la L_{space}(T_0) \ra$ for the 
spatial volume at time $T_0 < T$, even at the discretized level.
The details are given in the appendix.

\section{Topology changes}\label{topology}

In our non-perturbative regularization of 2d quantum gravity we have
so far not included the possibility of topology changes of space.
At the same time we found disagreement with the theory 
of Euclidean two-dimensional quantum gravity, whose properties 
we summarized in the introduction. We will now show that 
{\em if} one allows for spatial topology changes, one is led in
an essentially unambiguous manner to the theory of two-dimensional 
quantum gravity, 
as defined by dynamical triangulations or Liouville theory.

By a topology change of {\em space} in our Lorentzian setting
we have in mind the following: a baby universe may
branch off at some time $T$ and develop in the future, where it 
will eventually disappear in the vacuum, but it is
not allowed to rejoin the ``parent'' universe and thus change the 
overall topology of the two-dimensional manifold. This is 
a restriction we impose to be able to compare with 
the analogous calculation in usual 2d Euclidean quantum gravity.

It is well-known that such a branching leads to additional
complications, compared with the Euclidean situation, in the sense
that, in general, no continuum Lorentzian metrics which are smooth and 
non-degenerate everywhere can be defined on such space-times (see, for
example, \cite{ls} and references therein). 
These considerations do not affect the cosmological term in the
action, but lead potentially to contributions from the Einstein-Hilbert
term at the singular points where a branching or pinching occurs.

We have so far ignored the curvature term in the action since it
gives merely a constant contribution in the absence of topology
change. We will continue to do so in the slightly generalized
setting just introduced. The continuum results of \cite{ls} 
suggest that the contributions from the two singular points 
associated with each branching of a baby universe (one at the
branching point and one at the tip of the baby universe where it
contracts to a point) cancel in the action.
The physical geometry of these configurations may 
seem slightly contrived, but they may well be 
important in the quantum theory of gravity and deserve further 
study. However, for the moment our main motivation for introducing them
is to make contact with the 
usual non-perturbative Euclidean path-integral results.

We will use the rest of this section to demonstrate the following:
once we allow for spatial topology changes, 
\begin{itemize}
\item[(1)]this process completely dominates and changes the critical 
behaviour of the discretized theory, and
\item[(2)] the disc amplitude $W_\L(L)$ (the Hartle-Hawking wave function)
is uniquely determined, almost without any calculations.
\end{itemize}
Our starting point will be the discretized model introduced in
sec.\ \ref{model}.
Its disc amplitudes will be denoted by $w^{(b)}(l,g)$
and $w^{(b)}(x,g)$, where the superscript $^{(b)}$ indicates the 
``bare'' model without spatial topology changes.
Similarly, the transfer matrices will be labelled by 
$G_\l^{(b)}(l_1,l_2;t\equ 1)$ 
and $G_\l(l_1,l_2;t\equ 1)$, and the continuum amplitudes by
$W^{(b)}_\L (L)$, $W^{(b)}_\L(X)$.

\begin{figure}
\centerline{\hbox{\psfig{figure=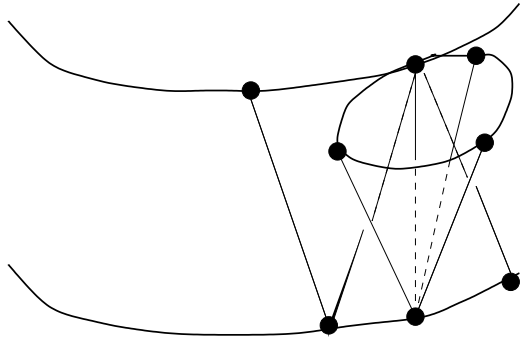,height=4.5cm,angle=0}}}
\caption[branching]{A ``baby universe'' branches off locally in 
one time-step.}
\label{branching}
\end{figure}

\begin{figure}
\centerline{\hbox{\psfig{figure=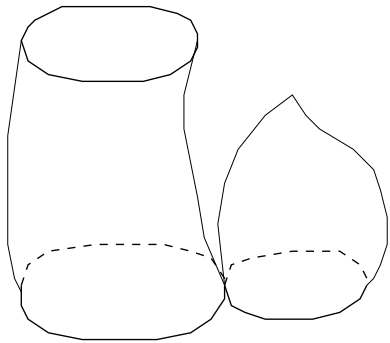,height=4.5cm,angle=0}}}
\caption[topchange]{An alternative representation of the process
in fig.\ \ref{branching}:
a ``baby universe'' is created by a global pinching.}
\label{topchange}
\end{figure}

There are a number of ways to implement the creation of baby universes, 
some more natural than others, but they all agree in the continuum limit,
as will be clear from the general arguments provided below. 
We mention just two ways of implementing such a change. The first
is a simple generalization of the forward step we have used
in the original model, where each vertex at time $t$ could connect to $n$
vertices at time $t+1$. We now allow in addition that these 
sets of $n$ vertices (for $n >2$) may form a baby universe with
closed spatial topology $S^1$, branching off from the rest.
The process is illustrated in fig.\ \ref{branching}.
An alternative and technically somewhat simpler way to implement the 
topology 
change is shown in fig.\ \ref{topchange}: stepping forward from 
$t$ to $t+1$ from a loop of length $l_1$ we create a
baby universe of length $l < l_1$ by pinching it off non-locally
from the main branch. We have checked that the 
continuum limit is the same in both cases. For simplicity we only 
present the derivation in the latter case.

Accounting for the new possibilities of evolution in each step
according to fig.\ \ref{topchange}, the new and old transfer matrices
are related by 
\beq{top1}
G_\l(l_1,l_2;1) = G_\l^{(b)}(l_1,l_2;1)+ 
\sum_{l=1}^{l_1-1} l_1 w(l_1\mi l,g)\,G_\l^{(b)}(l,l_2;1).
\eeq
The factor $l_1$ in the sum comes from the fact that the 
``pinching'' shown in fig.\ \ref{topchange} can take place at any of the 
$l_1$ vertices. As before, the new transfer matrix leads to new amplitudes
$G_\l(l_1,l_2;t)$, satisfying
\beq{top2a}
G_\l(l_1,l_2;t_1+t_2) = \sum_l G_\l(l_1,l;t_1)G_\l(l,l_2;t_2),
\eeq
and in particular 
\beq{top2}
G_\l(l_1,l_2;t) = \sum_l G_\l(l_1,l;1)\;G_\l(l,l_2;t\mi 1).
\eeq
Performing a (discrete) Laplace transformation of eq.\ \rf{top2}
leads to
\beq{top3}
G(x,y;g;t) = 
\ointz \left[G_\l^{(b)}(x,z^{-1};1)+ x \frac{\prt}{\prt x} 
\Bigl( w(x;g) G_\l^{(b)}(x,z^{-1};1)\Bigr) \right]  G(z,y;g;t \mi 1), 
\eeq 
or, using the explicit form of the transfer matrix $G_\l^{(b)}(x,z;1)$,
formula \rf{12},
\beq{top4}
G(x,y;g;t) = \Bigl[1+x\frac{\prt w(x,g)}{\prt x}
+ x w(x,g)\frac{\prt}{\prt x} \Bigr]
\, \frac{gx}{1-gx} \, G\Bigl( \frac{g}{1\mi gx},y;g;t\mi 1\Bigr).
\eeq
At this point neither the disc amplitude $w(x,g)$ 
nor the amplitude $G(x,y;g;t)$ are known. We will now 
show that they are uniquely determined if we assume 
that the boundary length scales canonically with the lattice
spacing, $L=a \, l$, implying a renormalized boundary 
cosmological constant $X$ with the dimension of mass, $x = x_c(1-aX)$.
In addition we assume that 
the dimension of the renormalized cosmological constant 
$\L$ is canonical too, $g=g_c(1-\oh \L a^2)$. Somewhat related 
arguments have been presented in different settings in \cite{ik,watabiki}.

It follows from relation \rf{top2a} that we need 
\beq{top7}
G_\l(l_1,l_2,t) \buildrel{a\rightarrow
0}\over\longrightarrow  a\, G_\L(L_1,L_2;T).
\eeq 
It is important for the following discussion that $G_\l(l_1,l_2;t)$
cannot contain a non-scaling part since from first principles (subadditivity) 
it has to decay exponentially in $t$.
By a Laplace transformation, using $x=x_c(1-a X)$ in the 
scaling limit, we thus conclude that 
\beq{top8}
G_\l(x,l_2,t) \buildrel{a\rightarrow
0}\over\longrightarrow G_\L(X,L_2,T),
\eeq
and further, by a Laplace transformation in $L_2$,  
\beq{top8x}
G_\l(x,y;t) \buildrel{a\rightarrow
0}\over\longrightarrow a^{-1} G_\L(X,Y;T).
\eeq

We will now show that the scaling of $w(x,g)$ is quite restricted 
too. The starting point is a combinatorial identity which the 
disc amplitude has to satisfy. The arguments are valid both 
for the disc amplitude in Euclidean quantum gravity and the 
disc amplitude we have introduced for our model in 
\rf{extra1}. The discretized version of formula \rf{extra1} is 
\beq{topz1}
w^{(b)}(x,g) := \sum_t G^{(b)}(x,l_2\equ 1;g;t) =G^{(b)}(x,l_2\equ 1;g) .
\eeq
It follows from eq.\ \rf{top8} that 
\beq{topz2}
w^{(b)}(x,g) \to a^{-1} W^{(b)}_\L (X).
\eeq
This scaling is indeed very different from the 
scaling of the disc amplitude in Euclidean 2d gravity where one has
\beq{topz3}
w^{(eu)}(x,g) = w_{ns}^{(eu)}(x,g)+a^{3/2}W^{(eu)}_\L (X).
\eeq
In relation \rf{topz3}, $w^{(eu)}_{ns}(x,g)$ is the non-scaling, 
analytic part of $w^{(eu)}(x,g)$,
and $W^{(eu)}_\L(X)$ is given by \rf{ob3}.
We will assume the general form 
\beq{an3}
w(x,g)= w_{ns}(x,g)+ a^{\eta}W_\L(X) + \mbox{less singular terms}
\eeq
for the disc amplitude. In the case $\eta < 0$ the first term is 
considered absent (or irrelevant). 
However, if $\eta >0$  a term like $w_{ns}$ will generically 
be present, since any slight redefinition of coupling constants of the 
model will produce such a term if it was not there from the beginning.

\begin{figure}
\centerline{\hbox{\psfig{figure=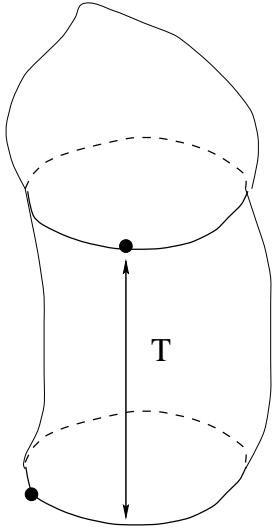,height=7cm,angle=0}}}
\caption[identity]{Marking a vertex in the bulk of $W_\L(X)$. The mark
has a distance $T$ from the boundary loop, which itself has one marked vertex.}
\label{identity}
\end{figure}

We will introduce an explicit mark in the bulk of $w(x,g)$
by differentiating with respect to $g$. This leads to the 
combinatorial identity 
\beq{an1}
g\ \frac{\prt w(x,g)}{\prt g} =
\sum_t \sum_l G(x,l;g;t) \, l\, w(l,g),
\eeq
or, after the usual Laplace transform,
\beq{an2}
g\ \frac{\prt w(x,g)}{\prt g}=
\sum_t \ointz G(x,z^{-1};g;t)\; \frac{\prt w(z,g)}{\prt z}.
\eeq 
The situation is illustrated in fig.\ \ref{identity}. 
A given mark has a 
distance $t$ ($T$ in the continuum) to the entrance loop. 
In the figure we have drawn all points which have the same distance 
to the entrance loop and which form a connected loop. In the 
bare model these are all the points at distance $t$. 
In the case where baby universes are allowed (which we have not
included in the figure),
there can be many disconnected loops at the same distance.
Let us assume a general scaling 
\beq{an3a}
T = a^{\ep} t,~~~~~\ep >0,
\eeq
for the time variable $T$ in the continuum limit. Above we saw that 
the bare model without baby universe creation corresponded to $\ep=1$. 
With the generalization \rf{an3a} we account for the fact that
by allowing for baby universes
we have introduced an explicit asymmetry between 
the time- and space-directions. 

Inserting \rf{an3} and \rf{an3a} into eq.\ \rf{an2} we obtain
\beq{an4}
\frac{\prt w_{ns}}{\prt g}- 2a^{\eta-2} 
\frac{\prt W_\L(X)}{\prt \L} = \frac{1}{a^{\ep}} 
  \int \d T \int dZ\;  G_\L(X,-Z;T) 
\bigg[ \frac{\prt w_{ns}}{\prt z} -a^{\eta-1}
\frac{1}{z_{c}} \frac{\prt W_\L(Z)}{\prt Z}
\bigg],
\eeq
where $(x,g)=(x_c,g_c)$ in the non-singular part. 

From eq.\ \rf{an4} and the requirement $\epsilon >0$ it follows that the 
only consistent choices for $\eta$ are
\begin{enumerate}
\item $\eta < 0$, i.e.\
\beq{an5c}
a^{\eta-2} \frac{\prt W_\L(X)}{\prt \L} = \frac{a^{\eta-1}}{2 a^{\ep}} 
  \int \d T \int dZ\;  G_\L(X,-Z;T) \;\frac{1}{z_{c}}
  \frac{\prt W_\L(Z)}{\prt Z},
\eeq
in which case we get $\ep =1$; and
\item $1< \eta < 2$. Here formula \rf{an4} splits into the two 
equations 
\beq{an5}
-a^{\eta-2} \frac{\prt W_\L(X)}{\prt \L} = \frac{1}{2 a^{\ep}}\, 
\frac{\prt w_{ns}}{\prt z}\bigg|_{z=x_c}\;
  \int \d T \int dZ\;  G_\L(X,-Z;T), 
\eeq  
and 
\beq{an6}
\frac{\prt w_{ns}}{\prt g}\bigg|_{g=g_c} = -\frac{a^{\eta-1}}{a^{\ep}}  
  \int \d T \int dZ\;  G_\L(X,-Z;T) 
\;\frac{1}{z_{c}} \frac{\prt W_\L(Z)}{\prt Z}.
\eeq
We are led to the conclusion that $\ep= 1/2$ and $\eta=3/2$, which are
precisely the values found in Euclidean 2d gravity.
Let us further remark that eq.\ \rf{an5} in this case becomes
\beq{an5a}
-\frac{\prt W_\L(X)}{\prt \L} = \mbox{const.}\;G_\L(X,L_2=0),
\eeq
which differs from \rf{extra1}, but agrees with \rf{extra2}.
Finally, eq.\ \rf{an6} becomes
\beq{an7}
\int \d T \int dZ\;  G_\L(X,-Z;T) 
\;\frac{\prt W_\L(Z)}{\prt Z} = \mbox{const.},
\eeq
which will be satisfied automatically if $\eta =3/2$ and $\ep=1/2$,
as we will show below.  
\end{enumerate}

At first sight it appears surprising that we cannot have values $\eta > 2$, 
since it is known that the so-called multi-critical
matrix models \cite{kazakov} have $\eta = m-1/2$, $m >2$ ($m=2$ 
corresponds to pure gravity). 
However, in these situations a generic coupling constant $g$
does {\em not} correspond to a cosmological constant, and
differentiation with respect to $g$ has a different meaning. 
The scaling in these theories is therefore different (they
describe non-unitary matter coupled to 2d quantum gravity and have 
negative-dimensional operators which dominate over the cosmological 
constant term). 

We will now analyze a possible scaling limit of \rf{top4}, 
assuming the canonical scaling 
$x=x_c(1-aX)$ and $g=g_c(1-\oh \L a^2)$.
In order that the equation have a scaling limit at all, $x_c,\, g_c$ and
$w_{ns}(x_c,g_c)$ must satisfy two relations which can be determined
straightforwardly from \rf{top4}. The remaining continuum equation reads
\bea
a^\ep\frac{\prt}{\prt T}\, G_\L(X,Y;T)& =&
-a \, \frac{\prt}{\prt X} \Bigl[ (X^2-\L) G_\L(X,Y;T)\Bigr] \nn
&&-a^{\eta-1}\frac{\prt}{\prt X} \Bigl[W_\L(X) G_\L(X,Y;T)\Bigr].
\label{top13}
\eea
The first term on the right-hand side of eq.\ \rf{top13} 
is precisely the one we have already encountered 
in our original model, while the second term 
is due to the creation of baby universes. 
Clearly the case $\eta < 0$ (in fact $\eta \leq 1$) is inconsistent
with the presence of the second term, i.e.\ the creation of baby universes. 
However, since $\eta <2$, 
the last term on the right-hand side of \rf{top13} will always 
dominate over the first term. {\em Once we 
allow for the creation of baby universes, this process will completely
dominate the continuum limit.} In addition we get $\ep = \eta-1$,
in agreement with \rf{an6}. It follows that $\eta > 1$ and 
we conclude that $\ep=1/2,\eta=3/2$ 
are the only possible scaling exponents if we allow for the creation 
of baby universes. 
These are precisely the scaling exponents obtained from 
two-dimensional Euclidean gravity in
terms of dynamical triangulations, as we have already remarked. The topology 
changes of space have induced an anomalous dimension for $T$. 
If the second term on the right-hand side of \rf{top13} had been absent,
this would have led to $\ep =1$, and the time $T$ scaling in the 
same way as the spatial length $L$.

In summary, in the case $(\eta,\ep)=(3/2,1/2)$ eq.\ \rf{top13} leads 
to the continuum equation
\beq{top16}
\frac{\prt}{\prt T}\, G_\L(X,Y;T) =
-\frac{\prt}{\prt X} \Bigl[W_\L(X) G_\L(X,Y;T)\Bigr],
\eeq 
which, combined with eq.\ \rf{an5a}, determines the continuum disc
amplitude $W_\L(X)$.
Integrating \rf{top16} with respect to $T$ and using that 
$G_\L(L_1,L_2;T\equ 0)=\delta(L_1\mi L_2)$, i.e.\
\beq{top18a} 
G_\L(X,L_2\equ 0;T\equ 0)=1,
\eeq
we obtain
\beq{top18}
-1 = \frac{\prt}{\prt X}\bigg[ W_\L(X) \frac{\prt}{\prt\L} W_\L(X)
\bigg].
\eeq
Since $W_\L(X)$ has length dimension --3/2, i.e. 
$W_\L^2(X) = X^3 F(\SL/X)$, the general solution must be of the form 
\beq{top19}
W_\L(X) = \sqrt{-2\L X + b^2 X^3+ c^2 \L^{3/2}}.
\eeq
From the very origin of $W_\L(X)$ as the Laplace transform of a disc 
amplitude $W_\L(L)$ which is bounded, it follows that $W_\L(X)$ has 
no singularities or cuts for $\mbox{Re}\, X >0$. This requirement 
fixes the constants $b,c$ in \rf{top19} such that 
\beq{top20}
W_\L(X) = b \Big(X-\frac{\sqrt{2}}{b\,\sqrt{3}} \,\SL\Big)
\sqrt{X+ \frac{2\sqrt{2}}{b\,\sqrt{3}}\SL},
\eeq
where the constant $b$ is determined by the model-dependent 
constant in \rf{an5a}. 
This expression for the disc amplitude agrees after a rescaling of 
the cosmological constant with $W^{(eu)}_\L (X)$ from 2d Euclidean 
quantum gravity. With $W_\L(X)$ substituted into
\rf{top16}, the resulting equation is familiar from the usual
theory of 2d Euclidean quantum gravity, where it has been
derived in various ways \cite{kkmw,ik,watabiki},
with $T$ playing the role of {\em geodesic distance} between 
the initial and  final loop. In particular, the 
intrinsic Hausdorff dimension is $d_H=4$ as soon as we allow 
for baby universe creation.

Let us finally comment on the difference between the equations for
the amplitudes \rf{an5c}-\rf{an6} for 
$(\eta,\ep) =(-1,1)$ and $(\eta,\ep)=(3/2,1/2)$ respectively. 
In the first case there are no baby universes and 
eq.\ \rf{an5c} entails that only {\em macroscopic loops} at a 
distance $T$ from the entrance loop are important (as illustrated 
by fig.\ \ref{identity}). On the other hand, the term $\prt W_\L(Z)/\prt Z$
which describes the presence of these macroscopic loops is absent in
eq.\ \rf{an5}.
This is consistent with eq. \rf{an5a}, which shows explicitly 
that the length of the upper loop in fig.\ \ref{identity} remains at
the cut-off scale, i.e. it never becomes macroscopic.
This agrees with the dominance of baby universes: 
at any point in space-time the probability 
for creating a little ``tip'' of the size of the cut-off scale 
will dominate. At the same time the right-hand side of eq.\ \rf{an5c}, i.e.\ 
eq.\ \rf{an7} will play no role in the case $1 < \eta <2$, being simply 
equal to a constant. This latter property is 
satisfied automatically, as can be seen by using an equation 
analogous to \rf{top16} for the exit instead of the entrance loop. 
Thus eq.\ \rf{an7} becomes proportional to  
\beq{last}
\int_0^\infty \d T \frac{\prt}{\prt T} \;G(X,L_2 \equ 0;T) = 
\mbox{const.},
\eeq
proving our previous assertion.

\section{Conclusions}

We have tried to construct a non-perturbative model of 2d Lorentzian
quantum gravity for universes with cylindrical topology and compact
space-like slices. 
The regularization of the model was performed in the spirit of dynamical 
triangulations, with fixed edge lengths for each triangle.
To encode the light-cone structure of the Lorentzian geometries,
we have restricted the sum over states to metric configurations 
with a discrete causal structure. 
A class of such causal triangulations
was constructed by varying the connectivity according to the 
rules laid down in sec.\ \ref{model}. 

If we regard the edge lengths as invariant geodesic distances
on the piecewise linear manifold corresponding to the 
triangulation, each triangulation defines a geometry, i.e.\
an equivalence class of metrics, constructed according to
Regge's prescription \cite{regge}. 
In this way the class of triangulations we are
considering constitutes a grid in the set of all geometries  
allowing for a causal structure. 
As in the case of dynamical triangulations,
we conjecture that this subset becomes uniformly dense at a critical 
point, where a continuum limit can be taken. As a consequence, neither 
a gauge-fixing nor a Faddeev-Popov determinant are needed. 
We should 
stress that this is an assumption, as it is in the framework of 
dynamical triangulations. 
In this latter approach it is corroborated by the fact that its results
agree with the continuum theory, whenever they can be compared. 
Also in the present model the assumption seems justified,  
since we obtain agreement with continuum 
calculations, and, when allowing the creation of baby universes, with 
the dynamical triangulation results.

A continuum limit of our model exists 
if we permit an analytic continuation in the coupling constants.
Our results then agree with the formal 
continuum calculations performed in 
the so-called proper-time gauge \cite{nakayama}. However, they 
disagree with those of dynamical triangulations, even if 
we stay ``Euclidean'', i.e.\ do not continue the cosmological
constant $\L \to -i\tL$ back to its original ``Lorentzian'' value.
For example, we obtain $d_H=2$ for the intrinsic Hausdorff dimension,
which shows that the typical geometry of a 
configuration entering in the path integral is much 
smoother than in the usual Euclidean version of dynamical triangulations. 
We have therefore shown that there exists a consistent and non-trivial
theory of pure two-dimensional quantum gravity, which can be defined
as the continuum limit of a discrete path integral, and which does not
lie in the universality class of the usual Liouville gravity.
However, as discussed in sec.\ \ref{topology}, once we admit
baby universe creation such that the spatial topology 
can change, we are forced back into the universality class of field
theories represented by Euclidean 2d quantum gravity.

The comparison between the Lorentzian and Euclidean sectors of our
model is not without subtleties. One can define quantities in the
Lorentzian sector that rely on the distinction between
space- and time-like directions, and which do not possess an
immediate analogue in the Euclidean theory. 
A related observation is that from a purely Lorentzian
viewpoint, our extension to configurations with branching
baby universes is not forced upon us from first principles.
Apart from introducing points where a branch of the
universe may disappear into nothing, the two-loop propagator 
$G_\L(L_1,L_2;T)$ has a peculiar ``acausal" property. Although the
causal structures on the individual histories are still well defined
in the case of topology change, the baby universes that contribute
to the amplitude $G_\L(L_1,L_2;T)$ consist entirely of vertices
that do not lie in the past of any part of the central universe,
(except for a single point of zero volume).
Their branches can even extend to times $T'>T$.

We could of course equally well have worked in a ``time-reflected''
picture where the baby universes are branches coming from the 
past which join the central universe at some later stage. This would have
avoided the problem with the propagator, but we would have had to
allow for the spontaneous creation of baby universes. 
We do not think that the distinction matters for our present purposes,
or that these features are a reason for serious concern at this stage.
Our main motivation for the extension of our model was  to understand
which part of the construction needs to be modified in order 
to make contact with the usual
Euclidean results. Since the Lorentzian configurations possess an
additional, causal structure, one has to expect subtleties of the kind 
just mentioned. 

Taking the possibility of topology changes more seriously would
seem to require a {\em third quantized} version of gravity (equivalently,
a string field theory), in order to deal in a consistent 
way with the creation and annihilation of universes of length $L$.
In the context of (Euclidean) non-critical string field theory, 
much progress has been made, in particular for $c\equ 0$, that is, pure 
2d Euclidean quantum gravity \cite{ik,watabiki,ik1}. However, its relation  
with a Lorentzian theory is totally unclear. Is any such formulation 
consistent with a concept of causality? Can one implement a restriction 
to causal structures in the Lorentzian path integral? Will the  
pathologies of Lorentzian metrics at points of topology change \cite{ls} 
play an important role in the path integral?
(Unlike in the special case of baby universes, one does not in
general expect a cancellation among the action contributions coming from 
such curvature singularities.) Maybe one will be forced to 
include geometries of both Euclidean and Lorentzian signature in 
the path integral, as was suggested in \cite{jeff1}. This work was 
motivated by some subtleties associated 
with the analytic continuation in the proper time, similar to those
we have encountered in our transfer-matrix approach.

One may summarize our results as follows: the problem of defining a theory
of two-dimensional quantum gravity as the continuum limit of a discrete 
path integral does not have a unique answer. 
Addressing the problem in a Euclidean setting, one obtains Liouville
gravity. In a Lorentzian framework, summing only
over metrics with a causal structure, the simplest consistent model
leads to a different, inequivalent theory (even modulo any analytic
continuation). {\em If} one wants to obtain agreement between the
two, one {\em must} extend the Lorentzian model by allowing
for spatial topology changes. However, from a purely Lorentzian
point of view it is unclear whether or not such configurations
should be included in the path integral. 

What can we conclude from the present work for the case of
{\em real} four-dimensional Lorentzian gravity? One clearly can
formulate similar causal restrictions on dynamically triangulated
four-geometries, but we do not know whether this prescription alone
would make the path integral {\em sufficiently} Lorentzian. 
The presence of the Einstein-Hilbert term makes the issue
of analytic continuation more complicated. 
However, if our results in two dimensions are anything to go by,
summing over geometries with a causal structure could 
in principle lead to a drastic change of the results. It may be worthwhile
to test this idea by numerical simulations, which are well-developped
for the case of Euclidean geometries (see, for instance, \cite{dt4} and 
\cite{recent} for most recent developments).

\subsection*{Acknowledgements}

We would like to thank A. Bovier, J. Louko, J. Greensite, C. Kristjansen 
and M. Staudacher for discussion and comments. 
J.A. acknowledges the support of the Professor Visitante Iberdrola 
Program and the hospitality of the University of Barcelona, where part
of this work was done, and the support of the Danish National
Foundation of Fundamental Research.

\section*{Appendix}

In this appendix we  calculate the average size of the spatial universe at
time $T_1$, when the total time is given by $T_1+T_2$.
We will show that in the continuum limit the spatial slices
are genuinely extended with a spatial volume $\sim 1/\SL$. 
One can use various boundary conditions. The computation is
simplest when one fixes the boundary cosmological constants $X,Y$ 
and lets the boundary lengths fluctuate according
to the distribution dictated by the choice of the 
boundary cosmological constants. 
The calculation can be done even at a discretized level. 
It is helpful to use the properties
\beq{36}
A_{t_1}A_{t_2}-B_{t_1}B_{t_2} = (1-F^2) A_{t_1+t_2},~~~~
B_{t_1}B_{t_2}-C_{t_1}C_{t_2} = (1-F^2) C_{t_1+t_2}
\eeq
\beq{37}
B_{t_1}A_{t_2}-C_{t_1}B_{t_2} = A_{t_1}B_{t_2}-B_{t_1}C_{t_2} =
(1-F^2) B_{t_1+t_2}
\eeq
of the coefficients $A_t$, $B_t$ and $C_t$,
which follow directly from the fact that $G(x,y;g;t)$,
parametrized as in \rf{17} must satisfy the fundamental composition
law \rf{10}.

For purely aesthetic reasons, we will 
consider the symmetric situation where both the entrance and exit loops 
are marked. We calculate the average size $\la l(t_1) \ra_{x,y}$ 
of a spatial universe at time-slice $t_1$ as 
\beq{38}
\la l(t_1) \ra_{x,y} = \frac{1}{G^{(2)}(x,y;g;t_1+t_2)}
\ointz G(x,z^{-1};g;t_1)\; z \frac{d}{dz} G^{(2)}(z,y;g;t_2),
\eeq
where the contour encloses the poles in the $z$-plane of 
$G(x,z^{-1};g;t_1)$ but not of $G^{(2)}(z,y;g;t_2)$.
From \rf{18c} and \rf{36}-\rf{37} the result of the contour integral is
\beq{39}
\la l(t_1) \ra_{x,y} = \frac{1}{1-F^2} \;
\frac{(A_{t_1}-B_{t_1}x)(A_{t_2}-B_{t_2}y)+
(B_{t_1}-C_{t_1}x)(B_{t_2}-C_{t_2}y)}{A_{t_1+t_2}-B_{t_1+t_2}(x+y)+
C_{t_1+t_2}xy}.
\eeq
If $t_1$ and $t_2$ go to infinity and $a$ to zero 
in such a way that $T_1 $ and $T_2$ stay finite, 
at the critical point $g_{c}=\frac{1}{2}$, but with $x$ and $y$ not 
going to 1 (the case of microscopic boundaries), we have 
\beq{40}
\la l(t_1) \ra_{x,y} = \frac{1}{a \, \SL}\; 
\frac{(1-\e^{-2\SLT_1})(1-\e^{-2\SLT_2})}{1-\e^{-2\SL (T_1+T_2)}}.
\eeq
This shows that $\la L(T_1)\ra = \la a\, l(t_1) \ra_{x,y}$ possesses
a continuum limit, independent of the size of the microscopic boundaries 
and independent of $T_1$ and $T_2$ if they are sufficiently large.

From \rf{39} we can directly find the expression $\la L(T_1)\ra_{X,Y}$
in the case of macroscopic loops:
\beq{41}
\la L(T_1)\ra_{X,Y} = \frac{1}{\SL}
\frac{\Bigl[(\SL+X)+\e^{-2\SLT_1}(\SL-X)\Bigr]\Bigl[
(\SL+Y)+\e^{-2\SLT_2}(\SL-Y)\Bigr]}{(X+\SL)(Y+\SL)-\e^{-2\SL (T_1+T_2)}
(\SL-X)(\SL-Y)}.
\eeq
The same expression could have been obtained starting directly 
from the continuum expression \rf{26} or equivalently from
\beq{42}
G^{(2)}_\L (X,Y;T)= \frac{4\L\ \e^{-2\SLT}}{\Bigl[(X+\SL)(Y+\SL)-
\e^{-2\SLT}(\SL-X)(\SL-Y)\Bigr]^2},
\eeq
and by calculating 
\bea
\la L(T_1)\ra_{X,Y}& =& \frac{1}{G^{(2)}_\L (X,Y;T_1+T_2)}\;
\int_{-i\infty}^{i\infty} dZ \;G_\L(X,-Z;T_1) \frac{d}{dZ}
G^{(2)}_\L(Z,Y;T_2) \nonumber \\
&=&  \frac{1}{G^{(2)}_\L (X,Y;T_1+T_2)}\;
\int_{-i\infty}^{i\infty} dZ\; G^{(2)}_\L(X,-Z;T_1) G^{(2)}_\L(Z,Y;T_2) 
\label{43}
\eea

Let us finally compute $\la L(T_1)\ra_{L_1,L_2}$ with the 
boundary lengths $L_1$ and $L_2$ kept fixed. In 
this case there is no problem with the marking of boundary loops
since the factors of $L_1$ and $L_2$ cancel in the normalization. 
Thus
\beq{44}
\la L(T_1)\ra_{L_1,L_2} = 
\frac{1}{G_\L (L_1,L_2;T_1+T_2)}\;
\int_0^\infty dL\; G_\L (L_1,L;T_1)\; L\; G_\L (L,L_2;T_2)
\eeq
Rather surprisingly, the integration can be performed explicitly using 
\beq{45}
\int_0^\infty dx \; x^3 \, \e^{-ax^2} I_1(\b x)I_1(\g x) = 
\frac{1}{8 a^3} \, \e^{\frac{\b^2+\g^2}{4a}}\Bigl[ 
(\b^2+\g^2)I_1\Bigl(\frac{\b \g}{2a}\Bigr)+ 
2\b\g I_0\Bigl(\frac{\b \g}{2a}\Bigr)\Bigr],
\eeq
and we obtain (before normalization), with the notation $S_i = \sinh \SL T_i$
and $C_i= \cosh \SL T_i$, $i=1,2$,
\bea
\lefteqn{\la L(T_{1})\ra^{unnorm}_{L_1,L_2} = 
\frac{\sqrt{\L L_1L_2}}{ L_2 S_{1+2}^3}  \; 
\e^{-\SL(\frac{C_1}{S_1} L_1+
\frac{C_2}{S_2}L_2)}
\;\e^{\frac{\SL\Bigl(L_1 \frac{S_2}{S_1}+
L_2 \frac{S_1}{S_2}\Bigr)}{S_{1+2} }}}
\nonumber \\
&& \times  \left[ 
( L_1 S_2^2 +L_2 S_1^2) I_1\Bigl(\frac{2 \sqrt{\L L_1L_2}}{S_{1+2}}\Bigr)
+ 2 S_1S_2 \sqrt{L_1L_2} 
I_0\Bigl(\frac{2 \sqrt{\L L_1L_2}}{S_{1+2}}\Bigr) \right].
\label{46}
\eea
Dividing by the normalization factor we obtain
\beq{47}
\la L(T_1)\ra_{L_1,L_2} = 
\frac{1}{ S_{1+2}^2} 
\left[ (L_1S_2^2+L_2 S_1^2) + 
2S_1S_2 \sqrt{ L_1L_2} 
\frac{I_0\Bigl(\frac{2 \sqrt{\L L_1L_2}}{S_{1+2}}\Bigr)}{
 I_1\Bigl(\frac{2 \sqrt{\L L_1L_2}}{S_{1+2}}\Bigr)}\right].
\eeq

\end{document}